\documentclass[epj]{svjour}
\usepackage{graphics}
\usepackage{epsfig}
\usepackage[usenames]{color}

\begin{document}
\title{Dynamical Coupled-Channel Analysis at EBAC}
\author{
T. -S. H. Lee\inst{1,2}
}                     
\institute{
Physics Division, Argonne National Laboratory, Argonne, IL 60439 \and
Excited Baryon Analysis Center (EBAC), 
Thomas Jefferson National Accelerator Facility, Newport News, Va. 22901  }
\date{Received: date / Revised version: date}
%
\abstract{
The status of the dynamical coupled-channel analysis at EBAC is reported.
} 
\authorrunning{T.-s. H. Lee}
\titlerunning{Coupled-channel analysis}                             
\maketitle
\section{Introduction}
\label{intro}
In this contribution, we report on the 
dynamical coupled-channel analysis being pursued
 at the Excited Baryon Analysis Center (EBAC) of Jefferson Laboratory.
EBAC was established in January, 2006. Its objective is to extract the 
parameters associated with the excited states ($N^*$) of the nucleon from 
the world data of meson production reactions, and to also 
develop theoretical interpretations of the extracted $N^*$ parameters.
 
Since $N^*$ states are unstable, their structure  must couple
with the reaction channels in the meson production reactions. 
To determine correctly the spectrum of $N^*$ states, an analysis of
the meson production data must account for the 
coupled-channel unitary condition. 
The extracted $N^*$ parameters can be interpreted correctly
only when the reaction mechanisms
in the short-range region where we want to map out the
$N^*$ structure have been accounted for.
To meet these two crucial requirements,  the dynamical coupled-channel
model developed in Ref.\cite{msl} is being applied at EBAC to analyze
the data of $\pi$, $\pi\pi$, $\eta$, $K$, and $\omega$ production.
It involves sloving the following coupled integral equations
\begin{eqnarray}
& &T_{\alpha,\beta}(p_\alpha,p_\beta;E)= V_{\alpha,\beta}(p_\alpha,p\beta) 
\nonumber \\
& &+ \sum_{\gamma}
 \int_0^{\infty} d p'  V_{\alpha,\gamma}(p_\alpha, p' )
G_{\gamma}( p' ,E)
T_{\gamma,\beta}( p'  ,p_\beta,E)  \,,
\label{eq:teq}  \\
& &V_{\alpha,\beta}= v_{\alpha,\beta}+
\sum_{N^*}\frac{\Gamma^{\dagger}_{N^*,\alpha} \Gamma_{N^*,\beta}}
{E-M^*} \,,
\label{eq:veq}
\end{eqnarray}
where
$\alpha,\beta,\gamma = \gamma N, \pi N, \eta N, 
$and $\pi\pi N$ which has $\pi \Delta, \rho N, \sigma N$ resonant components, 
$G_\gamma (p,E)$ is the propagator of channel $\gamma$,
$v_{\alpha,\beta}$ is defined by meson-exchange mechanisms, and
$\Gamma_{N^*,\beta}$ is related to the  
quark-gluon sub-structure of $N^*$.
 At the present time, it is reasonable
to interpret
$\Gamma_{N^*,\beta}$ in terms of 
hadron structure calculations with effective degrees of freedom, such as
the constituent quark model\cite{capstick}
and the model\cite{roberts} based on Dyson-Schwinger Equations.
In the near future, one hopes to relate $\Gamma_{N^*,\beta}$ to
Lattice QCD (LQCD) calculations.

If we take the on-shell approximation, Eq.(\ref{eq:teq}) is reduced into
an algebraic form of  K-matrix 
models\cite{said,kmatrix-1,kmatrix-2,kmatrix-3}
\begin{eqnarray}
T^{k}_{\alpha,\beta}(p_\alpha,p_\beta,E) &=& 
\sum_{\gamma}
V_{\alpha,\gamma}(p_\alpha,p_\gamma) \nonumber \\
& &\times[\delta_{\alpha,\gamma} + i \rho(p_\gamma)
T^{k}_{\gamma,\beta}(p_\gamma,p_\beta,E)] \,,
\label{eq:k-matrix}
\end{eqnarray}
where $\rho(p_\gamma)$ is an appropriate phase space factor.
Qualitatively speaking, the use of the on-shell model
based on  
$T^{k}_{\alpha,\beta}(p_\alpha,p_\beta,E)$ 
of Eq.(\ref{eq:k-matrix}) is to avoid an explicit treatment of
the reaction mechanisms in short range region where we want to map out the
quark-gluon sub-structure of $N^*$ states. Thus the $N^*$ parameters
extracted by using Eq.(\ref{eq:teq}) can be more directly interpreted
in terms of the  quark-gluon sub-structure of $N^*$.

In section 2, we briefly review the dynamical coupled-channel model
of Ref.\cite{msl}. The recent results from EBAC are reported in
section 3. Section 4 gives an outlook of EBAC.

\section{Dynamical Coupled-Channel Model }
\label{sec:1}
Within the formulation of Ref.\cite{msl}, Eqs.(\ref{eq:teq})-(\ref{eq:veq})
are derived from the following Hamiltonian
\begin{eqnarray}
H &=& H_0 + H_I  \\
& & \nonumber \\
H_0 &=& \sum_{MB} [\sqrt{m_B^2+k^2}+\sqrt{m_M^2+p^2}]  \\
& &\nonumber \\
H_I &=& \Gamma_V + v_{22} + v_{23} + v_{33} 
\label{eq:hi}
\end{eqnarray}
\begin{figure}[h!]
\begin{center}
\epsfig{file=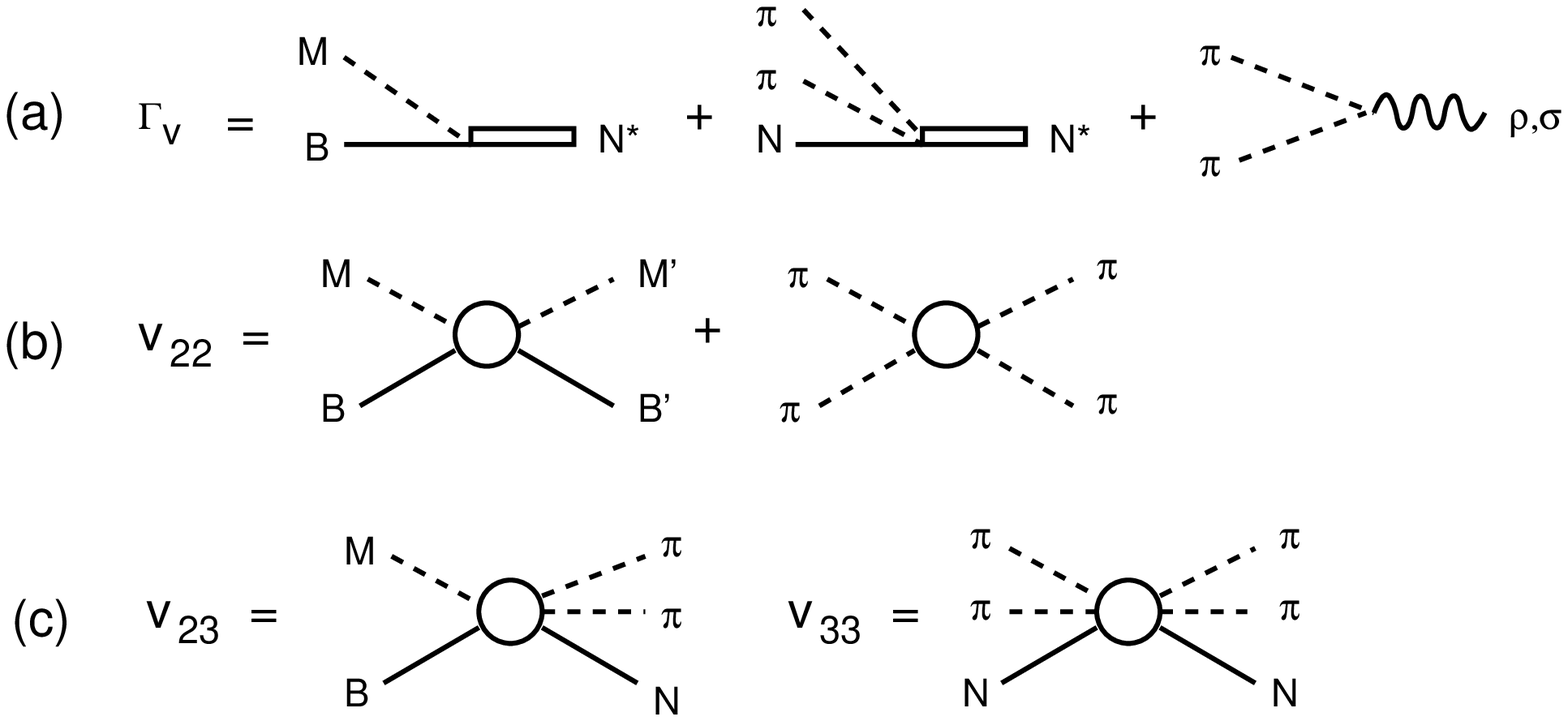,width=0.35\textwidth}
\end{center}
\caption{Basic mechanisms of the Model Hamiltonian defined in Eqs.(4)-(6). }
\label{fig:h}
\end{figure}

\begin{figure}[h!]
\begin{center}
\epsfig{file=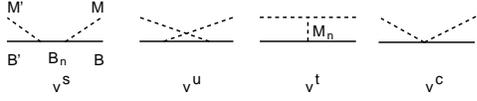,width=0.35\textwidth}
\end{center}
\caption{Meson-baryon (MB) interaction mechanisms of $v_{2,2}$ of Eq.(6).  }
\label{fig:mbmb}
\end{figure}

\begin{figure}[h!]
\begin{center}
\epsfig{file=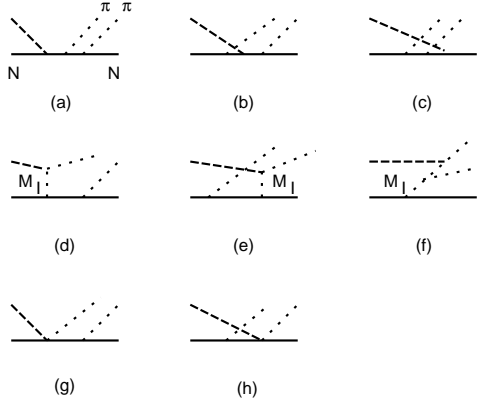,width=0.35\textwidth}
\end{center}
\caption{Examples of non-resonant mechanisms of $v_{MN,\pi\pi N}$
with $M = \pi$ or $ \gamma $ (denoted by long-dashed lines). $M_I$ denotes
the intermediate mesons ($\pi, \rho, \omega$).
}
\label{fig:v23}
\end{figure}
The non-resonant interactions $v_{22}, v_{23}, v_{33}$ are derived
from phenomenological Lagrangians by using an unitary transformation
method\cite{sl,sko}.
The interaction $v_{22}$ is defined by the
tree diagrams, as illustrated in
Fig.\ref{fig:mbmb}.
Examples of non-resonant mechanisms of
of $v_{23}$ are illustrated in Fig.\ref{fig:v23}.
The more complex $v_{33}$ is not included in the current analysis at EBAC.

\begin{figure}[h!]
\begin{center}
\epsfig{file=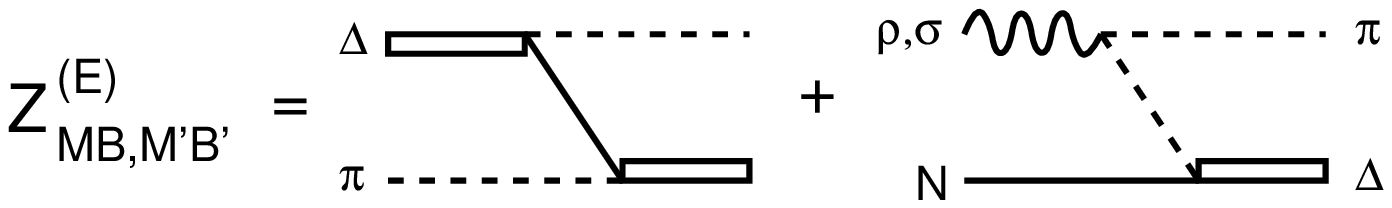,width=0.35\textwidth}
\caption{One-particle-exchange interactions
 $Z^{(E)}_{\pi\Delta,\pi\Delta}(E)$, $Z^{(E)}_{\rho N,\pi\Delta}$
and $Z^{(E)}_{\sigma N,\pi\Delta}$ of
Eq.(9).
}
\label{fig:z}
\end{center}
\end{figure}

\begin{figure}[h!]
\begin{center}
\epsfig{file=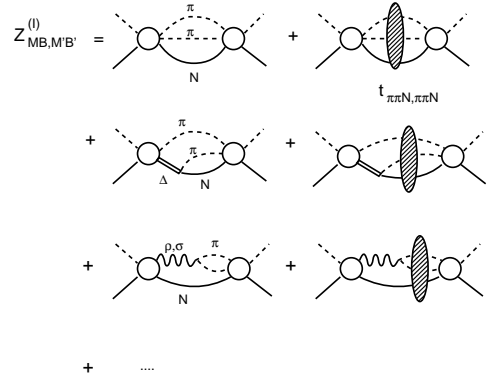,width=0.35\textwidth}
\caption{Examples of mechanisms included in $Z^{(I)}_{MB,M'B'}$ of
Eq.(9)}
\label{fig:zi}
\end{center}
\end{figure}

By using the projection operator  techniques, one can cast 
Eqs.(\ref{eq:teq})-(\ref{eq:veq}) for $2 \rightarrow 2$
meson-baryon ($MB$) transition amplitudes into the following form
\begin{eqnarray}
T_{a,b}(E) = t_{a,b}(E) + t^{R}_{a,b}(E)\,,
\label{eq:full-t}
\end{eqnarray}
where $a,b,c=\gamma N, \pi N, \eta N, \pi\Delta, \rho N, \sigma N$.
The non-resonant amplitude in Eq.(\ref{eq:full-t}) is defined by
\begin{eqnarray}
t_{a,b}(E)= V_{a,b} +\sum_{c} V_{a,c}G_c(E) t_{c,b}(E)\,,
\label{eq:nr-t}
\end{eqnarray}
where the driving term is
\begin{eqnarray}
V_{a,b}(E) = v_{a,b} +  Z^{(E)}_{a,b} (E) +Z^{(I)}_{a,b} (E).
\label{eq:veff}
\end{eqnarray}
The effects due to the three-body unitarity cuts are included
in $Z^{(E)}_{a,b} (E)$ and $Z^{(I)}_{a,b} (E)$, as illustrated in 
Figs.\ref{fig:z}-\ref{fig:zi}.


The resonant amplitude in Eq.(\ref{eq:full-t}) is defined by
\begin{eqnarray}
t^{R}_{a,b}(E) = \sum_{N^*_i, N^*_j}
\bar{\Gamma}^\dagger_{N^*_i, a}(E) [g(E)]_{i,j}
\bar{\Gamma}_{N^*_j, b}(E) \,,
\label{eq:r-t}
\end{eqnarray}
where the dressed vertex is
\begin{eqnarray}
\bar{\Gamma}_{N^*,a}(E) =
  { \Gamma_{N^*,a}} + \sum_{b} \Gamma_{N^*,b}
G_{b}(E)t_{b,a}(E)\,,
\label{eq:dress-v}
\end{eqnarray}
and the $N^*$ propagator is defined by
\begin{eqnarray}
g^{-1}_{i,j}(E)= E - M^0_i \delta_{i,j}
-\sum_{a}\bar{\Gamma}_{N^*_i,a}G_{a}(E)
{\Gamma}^\dagger_{N^*_j,a}\,.
\end{eqnarray}

Here we emphasize that the second term of the dressed
vertex Eq.(\ref{eq:dress-v}) is an necessary consequence
of the unitarity condition. This term describes the meson cloud
effects on the $N^*$ excitations. See Ref.\cite{jlss} for
a more detailed discussion on this point.

\begin{figure}[h!]
\begin{center}
\epsfig{file=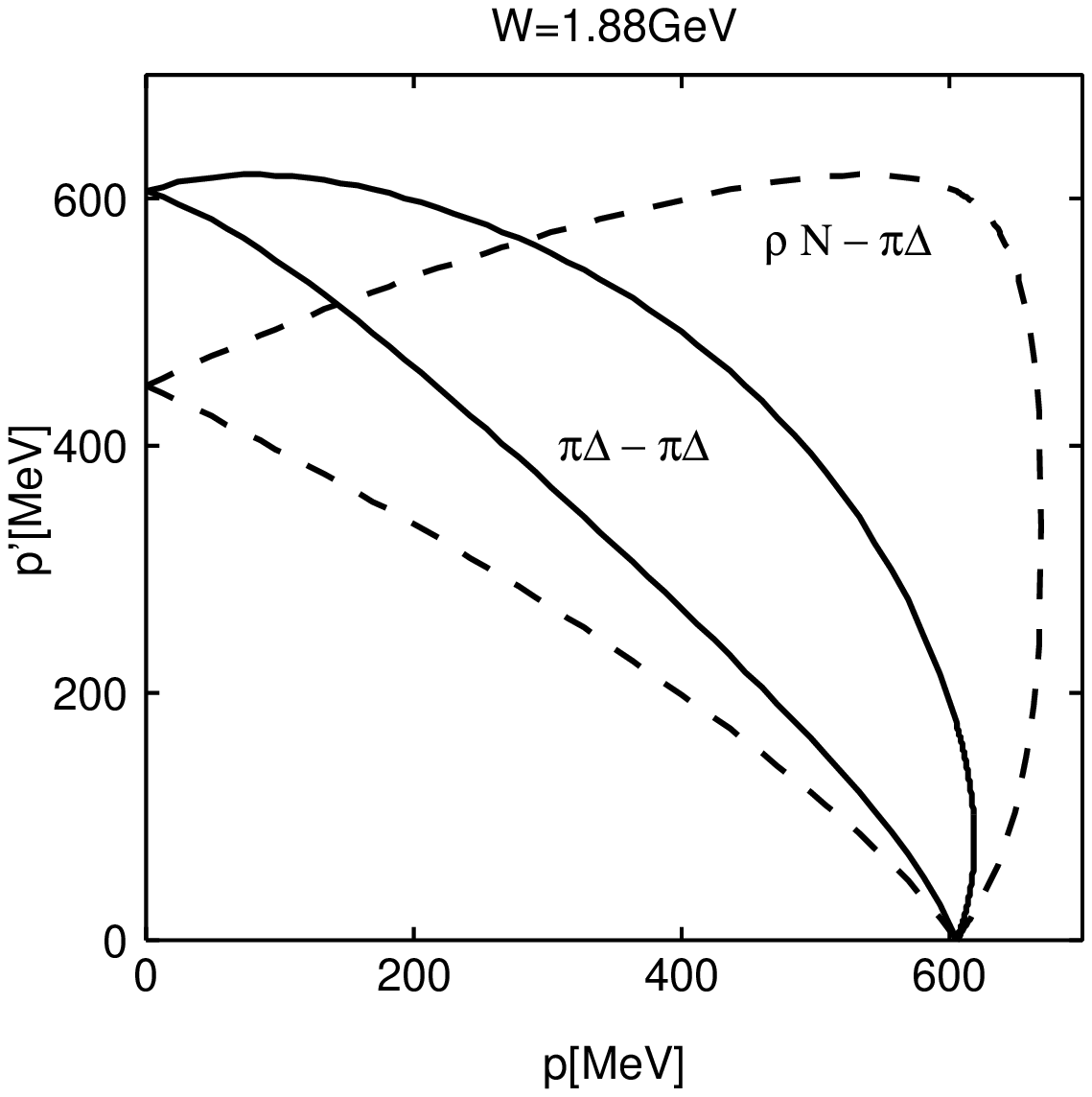,width=0.25\textwidth}
\end{center}
\caption{Logarithmically divergent moon-shape regions
of the matrix elements of
$Z^{(E)}_{\pi \Delta,\pi \Delta }(p',p,E)$ (solid curves) and
$Z^{(E)}_{\rho N,\pi \Delta }(p',p,E)$ (dashed curves).}
\label{fig:moon}
\end{figure}

\begin{figure}[h!]
\begin{center}
\epsfig{file=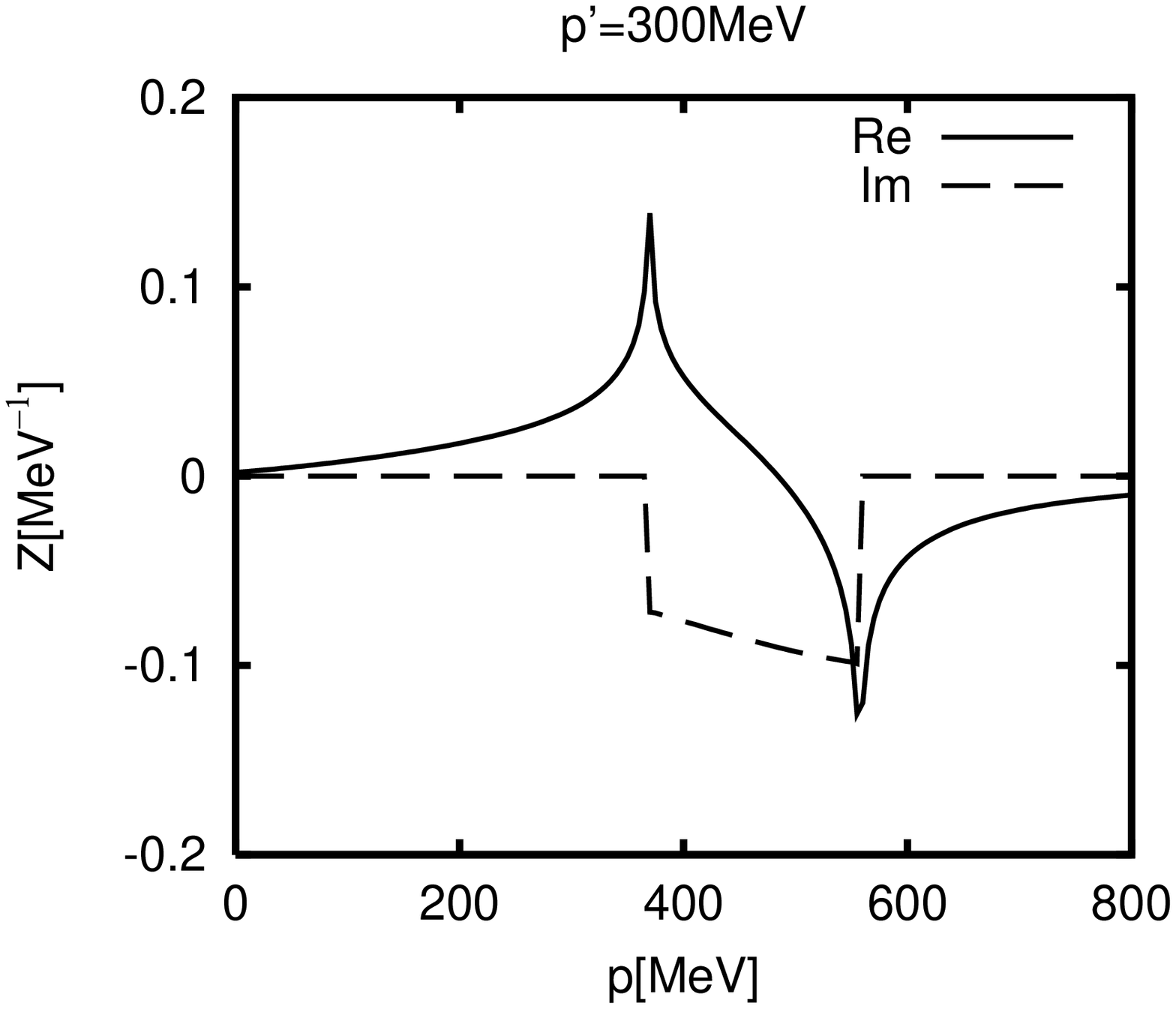,width=0.25\textwidth}
\end{center}
\caption{Matrix elements of the one-particle-exchange term
$Z^{(E)}_{\pi \Delta,\pi \Delta }(k,k',E)$
for $L=L'=1, J=5/2, T=1/2$ at $k'=300$ MeV/c and
$E=1.88$ GeV.}
\label{fig:z-pdpd}
\end{figure}

To solve Eq.(\ref{eq:nr-t}), we need to handle the singular structure
of $Z^{(E)}_{a,b}$ and $Z^{(I)}_{a,b}$. Their matrix elements diverge within
the moon-shapes region in Fig.\ref{fig:moon} and thus have
singular structure illustrated in Fig.\ref{fig:z-pdpd}.
We overcome this difficulty by using the Spline-function method
developed in Refs.\cite{AM-1,AM-2}.
 

\section{recent Results}
The computer code for solving the dynamical coupled-channel equations given in
section 2 has been developed. The numerical details, in particular
on the use of Spline-function methods, are explained 
in Ref.\cite{msl}. 

\begin{figure}[h!]
\begin{center}
\epsfig{file=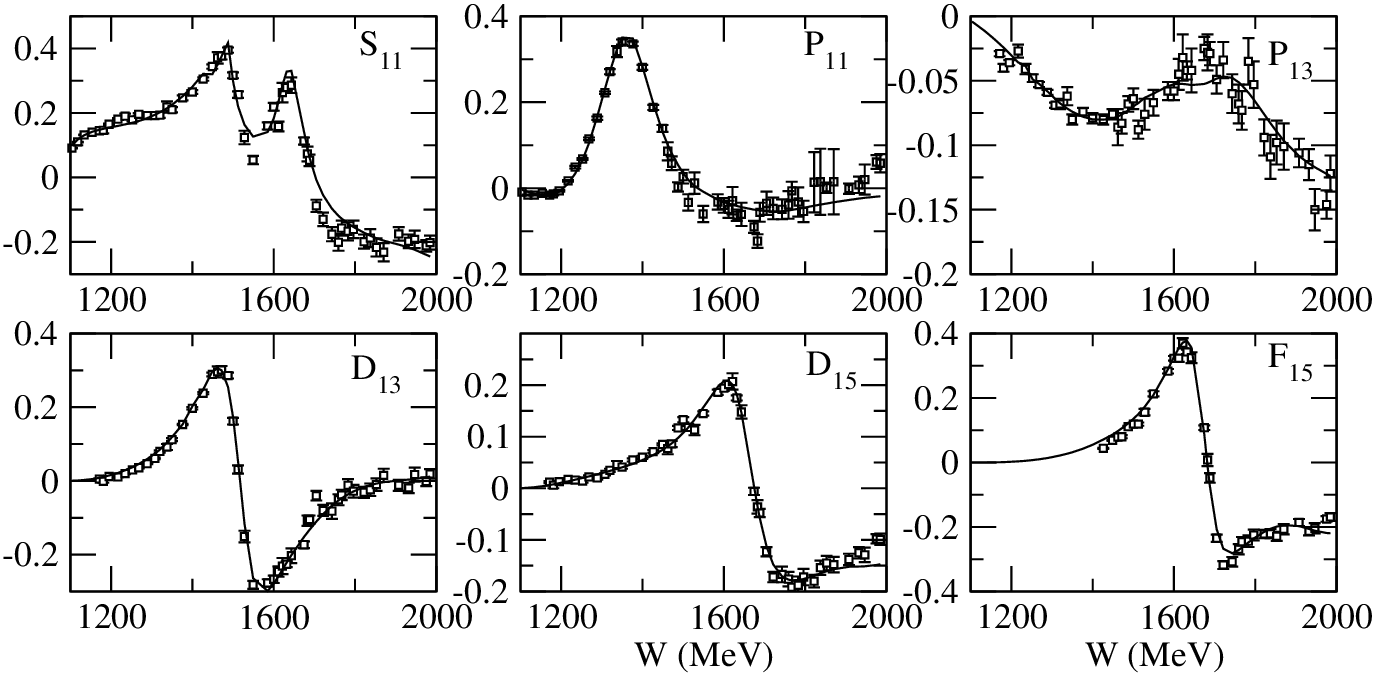,width=0.30\textwidth}
\caption{Fit to the $I=\frac{1}{2}$ $Re(T_{\pi N,\pi N})$  of
SAID\cite{said}. }
\label{fig:pin-1}
\end{center}
\end{figure}

\begin{figure}[h!]
\begin{center}
\epsfig{file=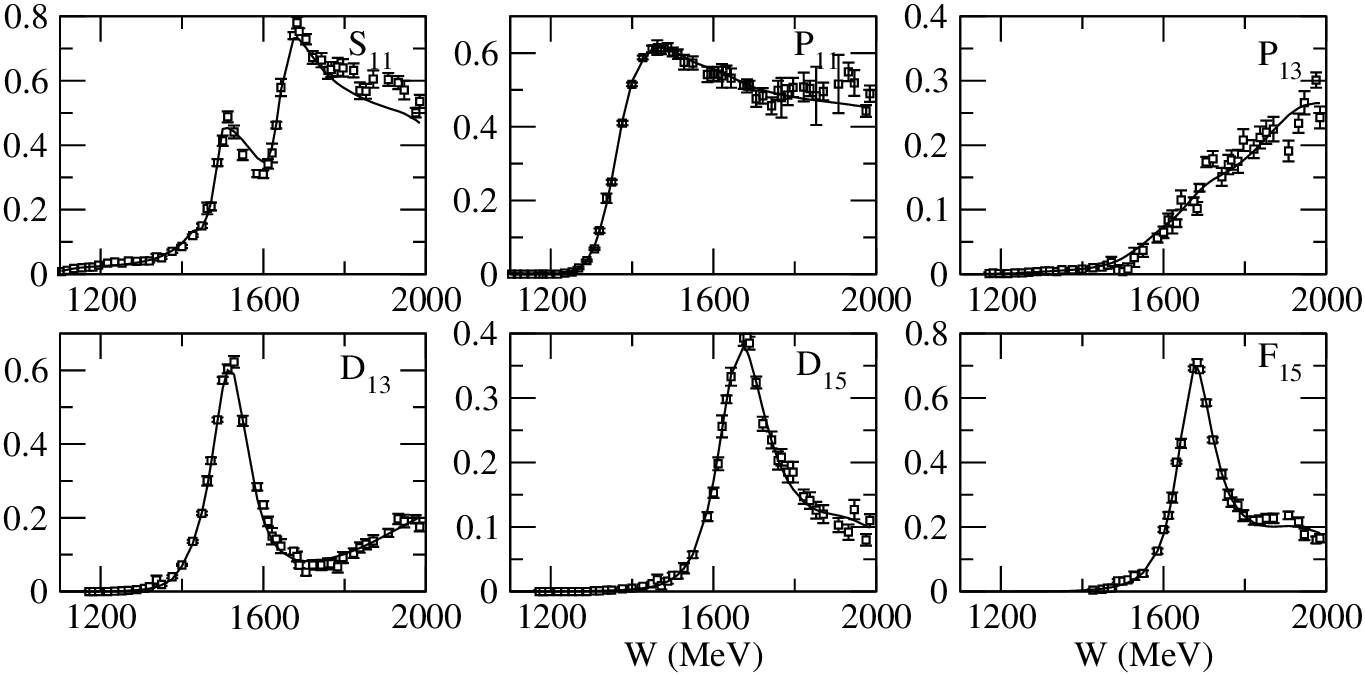,width=0.30\textwidth}
\caption{Fit to the $I=\frac{1}{2}$ $Im(T_{\pi N,\pi N})$  of
SAID\cite{said}, }
\label{fig:pin-2}
\end{center}
\end{figure}

Our first task is to
determine the parameters of
the hadronic interactions of
Hamiltonian defined by Eq.(\ref{eq:hi}).
This has been done\cite{jlms}
by fitting the $\pi N$ scattering data.
One or two bare $N^*$ 
states in each of $S$, $P$, $D$, and $F$ partial waves are included
to generate the resonant amplitudes, defined by Eq.(\ref{eq:r-t}), in the fits.
The parameters of the model are first determined by fitting as much as
 possible the empirical $\pi N$ elastic scattering amplitudes
of SAID\cite{said} up to 2 GeV. 
Some of our fits are shown in Figs.\ref{fig:pin-1}-\ref{fig:pin-2}.


\begin{figure}[h!]
\begin{center}
\epsfig{file=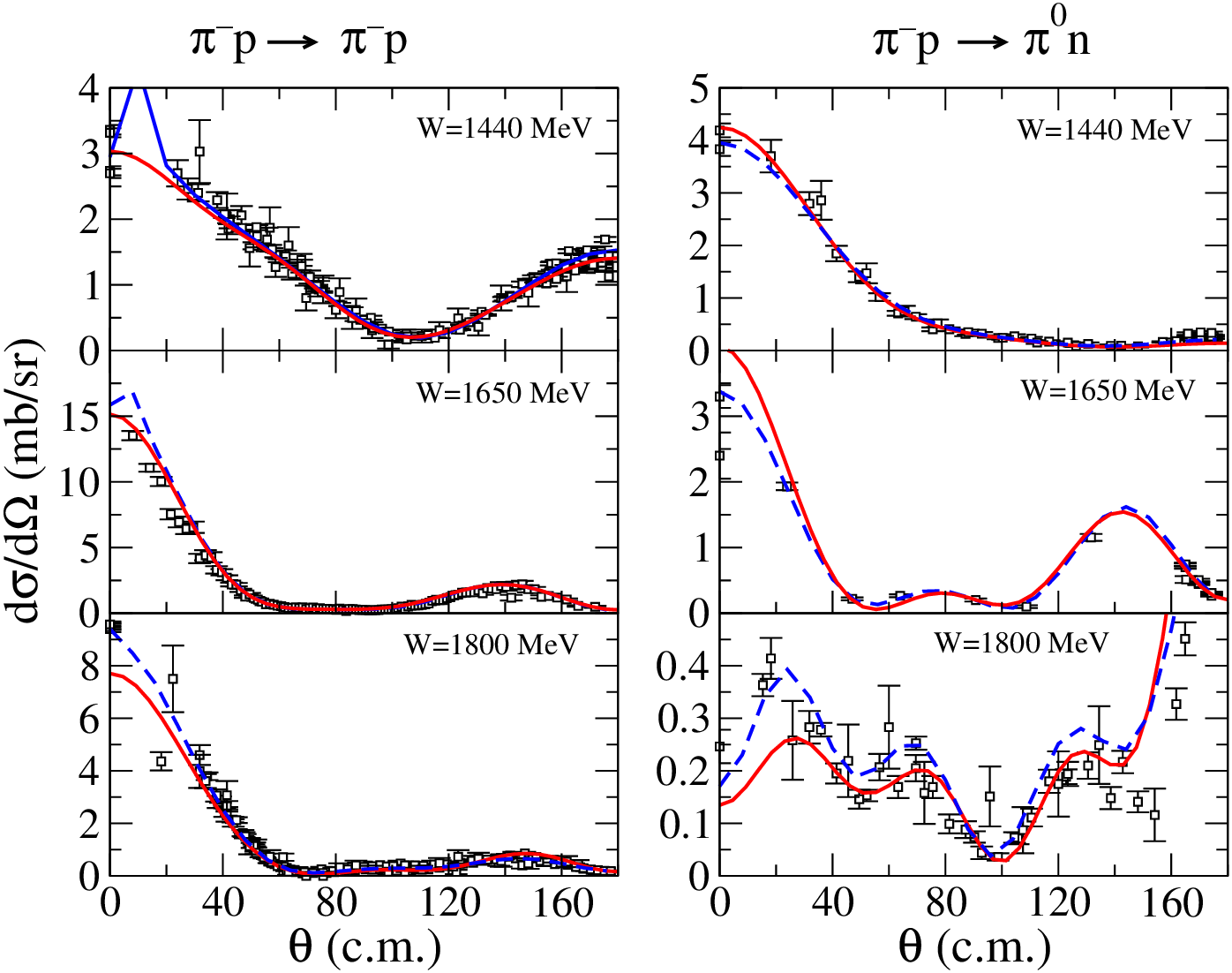,width=0.30\textwidth}
\caption{ Differential cross section for several
different center of mass energies. Solid curves correspond to our
model while blue dashed lines correspond to the SP06 solution of
SAID~\cite{said}. All data have been obtained through the SAID online
applications.~Ref.~\cite{said}.}
\label{fig:pin-dif}
\end{center}
\end{figure}
We then refine and confirm the resulting parameters by
directly comparing the predicted differential cross section and
target polarization asymmetry
with the original data of the elastic $\pi^{\pm} p \rightarrow \pi^{\pm} p$
and charge-exchange $\pi^- p \rightarrow \pi^0 n$ processes.
Typical results are shown in Fig.\ref{fig:pin-dif}.
The predicted total cross sections of $\pi N$ reactions
are also in good agreement
with the data, 
as shown in the left hand side of Fig.\ref{fig:pin-tot}.
Our predictions of the partial total cross sections to each
channel, as given in the right hand side of Fig.\ref{fig:pin-tot},
need to be refined  by also fitting 
 the $\pi N \rightarrow \pi\pi N$ data.
A combined fit to both the data of $\pi N$ elastic scattering 
and $\pi N \rightarrow \pi\pi N$ is now being pursued at EBAC.
Fig.\ref{fig:pipi} is a result from this very challenging
task, showing
the importance
of the $\pi\pi N$ cut in determining the predicted
invariant mass distribution of $\pi N \rightarrow \pi\pi N$ reaction.
Similar pronounced effects on 
$\gamma N \rightarrow \pi\pi N$ reactions 
have been presented in Ref.\cite{msl}.

\begin{figure}[h!]
\begin{center}
\epsfig{file=pimp-total.eps,width=0.35\textwidth}
\end{center}
\caption{Predicted $\pi^- p$ total cross sections
}
\label{fig:pin-tot}
\end{figure}

\begin{figure}[h!]
\begin{center}
\epsfig{file=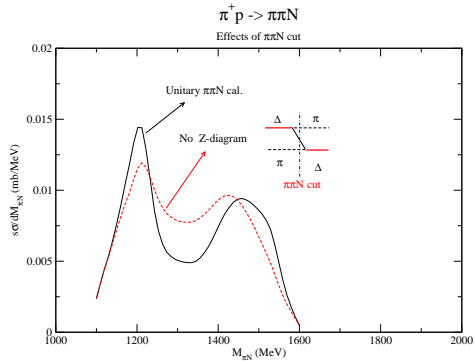,width=0.30\textwidth,angle=-90}
\end{center}
\caption{The $\pi N$ invariant mass
distribution of $\pi N \rightarrow \pi\pi N$ reaction at $W =$ 1880 MeV.
}
\label{fig:pipi}
\end{figure}

\begin{figure}[b!]
\begin{center}
\epsfig{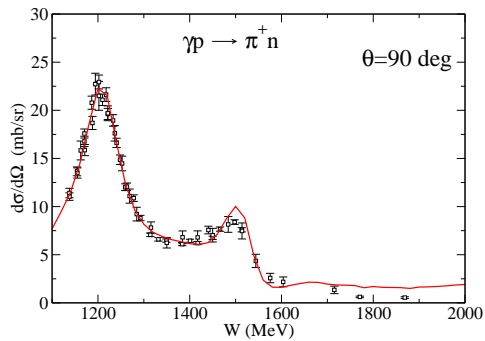}
\end{center}
\caption{Fit to the $\gamma p \rightarrow \pi^+ p$ up to $W=2$ GeV }
\label{fig:gpi}
\end{figure}

With the hadronic parameters determined, we are now moving to
analyzing the data of electromagnetic production of $\pi$ and $\pi\pi$.
Here the only parameters to be determined are the
bare $\gamma N \rightarrow N^*$ vertex interactions.
In Fig.\ref{fig:gpi}, we show the preliminary results for
$\gamma p \rightarrow \pi^+ n$. 
In Fig.14 we show the coupled-channel effects on the 
$\gamma N \rightarrow \pi N$ reactions
in the $\Delta$ excitation region.
More detailed results will be
presented in Ref.\cite{jlmss}.

\section{Outlook}

The dynamical coupled-channel model developed in Ref.\cite{msl} is being 
applied at EBAC to analyze the world data of meson production reactions.
The analysis of $\pi$ and $\pi\pi$ production data is proceeding well with
some encouraging preliminary results.
We have also started to analyze the data of
$\eta$, $K$, and $\omega$ production. Here the main task is to extend
the model  to include the
interactions associated with the $K \Lambda$,
$K \Sigma$, and $\omega N$ channels.  

In parallel, we are also developing an analytical continuation method to extract
the resonance poles and residues from the partial wave amplitudes
predicted by the constructed dynamical 
coupled-channel model. Furthermore, we are 
investigating the extent to which the  resonance poles and 
bare $N^*$ parameters extracted from our analysis can be related to the
current hadron structure calculations; in particular 
the Lattice QCD calculations.

\begin{figure}[h!]
\begin{center}
\epsfig{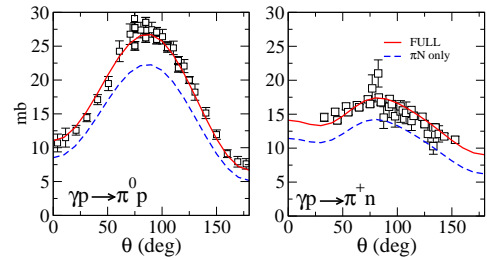}
\caption{
Coupled-channel effects on the
$\gamma N \rightarrow \pi N$ in the $\Delta$ region}
\end{center}
\label{fig:cc-d}
\end{figure}

This work is supported by the U.S. Department of Energy,
Office of Nuclear Physics Division, under contract No. DE-AC02-06CH11357,
and Contract No. DE-AC05-060R23177
under which Jefferson Science Associates operates Jefferson Lab.

%

\begin{thebibliography}{}
\bibitem{msl}
A. Matsuyama, T. Sato, T.-S. H. Lee, Phys. Rept. {\bf 439}, 193 (2007).
\bibitem{capstick}
See the review by 
S. Capstick S and W. Roberts, {\it Prog. Part. Nucl. Phys.} 
{\bf 45} S241 (2000)
\bibitem{roberts}
See the review by P. Maris and C.D. Roberts,
Int.J.Mod.Phys. {\bf E12} 297(2003).
\bibitem{said}
Arndt R, Strakovsky I, Workman R, 
 Int. J. Mod. Phys., {\bf A18} 449 (2003)

\bibitem{kmatrix-1}
D. Drechsel, O. Hanstein, S. Kamalov and L. Tiator   
Nucl. Phys. {\bf A645}, 145 (1999)
\bibitem{kmatrix-2}
I. Aznauryan, Phys. Rec. {\bf C71}, 01520 (2005)
\bibitem{kmatrix-3}
V. Shklyar, H. Lenske, U. Mosel and G. Penner,
Phys. Rev. {\bf C71}, 055206 (2005)


\bibitem{jlss}
B.~Julia-Diaz, T.-S. H. Lee, T.~Sato and L.~C.~Smith,
  Phys.\ Rev.\  C {\bf 75}, 015205 (2007).

\bibitem{sl}
T.~Sato and T.-S.~H. Lee, Phys. Rev. C {\bf 54}, 2660 (1996).
\bibitem{sko}
M. Kobayashi, T. Sato, and H. Ohtsubo, Prog.\ Theor.\ Phys.\  {\bf 98}, 927 (1997).
\bibitem{AM-1}
A. Matsuyama, Phys. Lett. B{\bf 152}, 42 (1984).

\bibitem{AM-2}
A. Matsuyama and T.-S. H. Lee, Phys. Rev. C {\bf 34}, 1900 (1986).

\bibitem{jlms}
B. Julia-Diaz, T.-S. H. Lee, A. Matsuyama, T. Sato, to appear
in Phys. Rev. C (2007).

\bibitem{jlmss}
B. Julia-Diaz, T.-S. H. Lee, A. Matsuyama, T. Sato, L.C. Smith,
in preparation

\end{thebibliography}
%

\end{document}